%% file: text.tex
\documentclass[graybox]{svmult}

\usepackage{type1cm}        
\usepackage{makeidx}         
\usepackage{graphicx}        
\usepackage{multicol}        
\usepackage[bottom]{footmisc}

\usepackage{newtxtext}       %
\usepackage[varvw]{newtxmath}       

\usepackage[tableposition=top,font=small,skip=5pt]{caption}

\makeindex     

\input{text/packages}

\begin{document}

\title*{Quantum Computing Methods for Malware Detection}
%

\author{Eliška Krátká\orcidID{0009-0000-5152-4670} and\\Aurél Gábor Gábris\orcidID{0000-0002-2671-6328}}
%
\institute{Eliška Krátká \at Faculty of Information Technology, Czech Technical University in Prague, Prague, Czechia, \email{kratkeli@fit.cvut.cz}
\and Aurél Gábor Gábris \at Faculty of Nuclear Sciences and Physical Engineering, Czech Technical University in Prague, Prague, Czechia,  \email{gabris.aurel@fjfi.cvut.cz}
}

\maketitle

\abstract*{In this paper, we explore the potential of quantum computing in enhancing malware detection through the application of Quantum Machine Learning (QML). Our main objective is to investigate the performance of the Quantum Support Vector Machine (QSVM) algorithm compared to SVM. A publicly available dataset containing raw binaries of Portable Executable (PE) files was used for the classification. The QSVM algorithm, incorporating quantum kernels through different feature maps, was implemented and evaluated on a local simulator within the Qiskit SDK and IBM quantum computers. Experimental results from simulators and quantum hardware provide insights into the behavior and performance of quantum computers, especially in handling large-scale computations for malware detection tasks. The work summarizes the practical experience with using quantum hardware via the Qiskit interfaces. We describe in detail the critical issues encountered, as well as the fixes that had to be developed and applied to the base code of the Qiskit Machine Learning library. These issues include missing transpilation of the circuits submitted to IBM Quantum systems and exceeding the maximum job size limit due to the submission of all the circuits in one job.}

\abstract{In this paper, we explore the potential of quantum computing in enhancing malware detection through the application of Quantum Machine Learning (QML). Our main objective is to investigate the performance of the Quantum Support Vector Machine (QSVM) algorithm compared to SVM. A publicly available dataset containing raw binaries of Portable Executable (PE) files was used for the classification. The QSVM algorithm, incorporating quantum kernels through different feature maps, was implemented and evaluated on a local simulator within the Qiskit SDK and IBM quantum computers. Experimental results from simulators and quantum hardware provide insights into the behavior and performance of quantum computers, especially in handling large-scale computations for malware detection tasks. The work summarizes the practical experience with using quantum hardware via the Qiskit interfaces. We describe in detail the critical issues encountered, as well as the fixes that had to be developed and applied to the base code of the Qiskit Machine Learning library. These issues include missing transpilation of the circuits submitted to IBM Quantum systems and exceeding the maximum job size limit due to the submission of all the circuits in one job.}

\input{text/sections/1-section}
\input{text/sections/2-section}
\input{text/sections/3-section}

\input{text/sections/4-section}

\input{text/sections/5-section}

\acknowledgement{This work was supported by the Grant Agency of the Czech Technical University in Prague, grant No. SGS23/211/OHK3/3T/18 funded by the MEYS of the Czech Republic.}

\bibliographystyle{plain}
\bibliography{references.bib}

\end{document}

%% file: text/sections/1-section.tex
\section{Introduction}\label{sect:1}
    Quantum computing has opened up new possibilities for addressing complex computational problems that classical computers struggle to solve. Quantum computers exploit the principles of quantum mechanics, such as superposition and entanglement, which allow them to perform parallel computations and potentially achieve exponential speedup for specific tasks.
    
    In recent years, a significant milestone in quantum computing has been the development of noisy intermediate-scale quantum (NISQ) devices~\cite{PreskillNISQ}. NISQ devices are the class of quantum computers characterized by their intermediate scale in the number of qubits. Unlike universal fault-tolerant quantum computers, which are still a theoretical goal, NISQ devices operate with a limited number of qubits and suffer from errors due to the noise in the quantum hardware~\cite{NISQ2022}. They typically have tens to hundreds of qubits, larger than what can be simulated classically but smaller than required for error correction and fault tolerance~\cite{NISQ2022}.
    
    One promising research area on the presently available NISQ computers is the combination of quantum computing and machine learning, known as Quantum Machine Learning (QML). Over the last decade, there have been significant advances in the QML field, including conventional machine learning algorithms that can be enhanced using quantum techniques and entirely new quantum machine learning algorithms explicitly designed to run on quantum computers~\cite{Gujju2023}.
    
    In this chapter, we explore the potential of applying QML to a practical problem from information security: malware detection. Malware detection is the process of identifying malicious software. This task is typically framed as a binary classification problem, where the goal is to distinguish between two categories: malicious and benign (harmless) software~\cite{shahzad2024}. Machine learning models are well-suited for solving this type of problem. Given the increasing volume and variety of new threats, malware detection based on machine learning has become a popular approach in modern antivirus programs~\cite{avastML, kasperskyML}.

    Our research focuses on the Quantum Support Vector Machine (QSVM) algorithm and its application to malware detection. A key part of the work involves running the QSVM on quantum computers from IBM. Executing the algorithm on real quantum hardware presents unique challenges, making the process more difficult than running the same calculations on quantum computer simulators.
    
    The QSVM algorithm combines the conventional Support Vector Machine (SVM) with a quantum kernel. We study and implement the quantum kernel using a quantum computer. The SVM model is then fitted with the precomputed quantum kernel and trained on a classical computer. We assess the performance of the QSVM in terms of the model's accuracy and compare its results to SVM using conventional kernels.

    We organize our work into three parts, with each covered in the following sections. In Section~\ref{sect:2}, we provide the necessary background on the quantum computing aspects of our research, explaining how quantum kernels in QSVM differ from conventional ones and how they are computed using quantum computers. We then introduce Qiskit~\cite{qiskit2024} and IBM Quantum~\cite{IBMQuantumPlatform}, highlighting their roles in implementing the algorithm and accessing quantum hardware. Section~\ref{sect:3} focuses on our implementation, emphasizing the challenges faced during the development process for quantum processors and how we addressed them. Section~\ref{sect:4} presents the performed experiments and the achieved results.
    

%% file: text/sections/2-section.tex
\section{Background}\label{sect:2}
    In this section, we explain the core concepts and principles underlying quantum computation, which are necessary to understand before we focus on the QSVM algorithm. We examine the theoretical foundations of QSVM, describe how it operates on quantum computers and the advantages it offers over its classical counterpart.

    Furthermore, we introduce the Qiskit and its machine learning module~\cite{qiskit2024, QiskitMachineLearning}. Through Qiskit, researchers can develop quantum algorithms and access quantum computers from IBM, which are available through the IBM Quantum platform~\cite{IBMQuantumPlatform}. We discuss the role of Qiskit in our research in implementing QSVM and performing quantum experiments on real quantum processors.

    \subsection{Terminology}\label{subsec:2.1}
    We follow the definitions and explanations of key terms laid down by Nielsen and Chuang in~\cite{NielsenChuang2010}. The only prerequisite is a basic understanding of elementary linear algebra and classical computing. 

    The standard notation for linear algebra in quantum mechanics and quantum computing is known as braket notation, which consists of two elements, bra and ket. The ket, written as~$\ket{\psi}$, denotes a vector in the vector space. The bra, written as~$\bra{\psi}$, represents a dual vector to the ket. An inner product of two vectors~$\ket{\psi}$ and~$\ket{\varphi}$ is denoted by~$\braket{\varphi}{\psi}$. The inner product is formally a map
    $$
        \langle\cdot, \cdot\rangle: V \times V \to \mathbb{C},
    $$
    where~$V$ is a vector space over~$\mathbb{C}$, which satisfies the following three properties for all vectors~$x,y,z \in V$ and all scalars~$\alpha \in \mathbb{C}$:
    \begin{align*}
        \begin{aligned}
            &\text{1.}\ \braket{x}{\alpha y + z} = \alpha \braket{x}{y} + \braket{x}{z} &&\text{\emph{(linearity in the second argument),}} \\
            &\text{2.}\ \braket{x}{y} = \braket{y}{x}^{*} &&\text{\emph{(conjugate symmetry),}} \\
            &\text{3.}\ \braket{x}{x} \geq 0\ \text{with equality if and only if}\ \ket{x} = 0 &&\text{\emph{(positive definiteness),}}
        \end{aligned}
    \end{align*}
    where * is a complex conjugate and 0 is a zero vector~\cite{Axler2024}. Quantum computing operates within a finite-dimensional Hilbert space, which in this context is equivalent to a complex vector space~$\mathbb{C}^n$ with the inner product.

    A quantum bit, known as a qubit, serves as the fundamental unit of information in quantum computing. While classical computing processes information using bits, which are binary variables capable of holding values 0 or 1, quantum computing utilizes qubits.
    
    A state of the qubit, the quantum state, is described by a unit vector in a two-dimensional Hilbert Space, which we further refer to as a quantum state space. The states~$\ket{0}$ and~$\ket{1}$ denote the fundamental computational basis states of the qubit, forming an orthonormal basis. Any quantum state of the qubit can be expressed as a linear combination of~$\ket{0}$ and~$\ket{1}$, meaning a qubit can exist in a superposition of these states. For example, the state
    $$
        \ket{\psi} = \alpha\ket{0} + \beta\ket{1},
    $$
    represents the qubit in the superposition of~$\ket{0}$ and~$\ket{1}$.
    
    The complex numbers~$\alpha$ and~$\beta$ referred to as probability amplitudes satisfy
    $$
        \lvert \alpha \rvert^2 + \lvert \beta \rvert^2 = 1.
    $$
    They encode the probability of each outcome and the associated phase information. In contrast to a classical probability distribution, which only considers the real numbers, probability amplitudes incorporate both magnitude and phase. The absolute squares of the probability amplitudes give the probabilities of the possible outcomes occurring when measured in the computational basis.
  
    Measurement plays an essential role in quantum computing. While the state of a classical bit can be observed without altering it, the qubit in superposition cannot be directly measured without affecting its quantum state. Upon measurement, the qubit \textit{collapses} into one of the basis states, giving an outcome of either~$\ket{0}$ with a probability of~$\lvert \alpha \rvert^2$ or~$\ket{1}$ with a probability of~$\lvert \beta \rvert^2$. Consequently, quantum states inherently embody non-determinism, as their measurement is probabilistic and fundamentally different from classical systems.

    The building blocks of quantum computing are quantum gates and circuits. Quantum gates are basic operations that manipulate qubits, similar to classical logic gates. They come in various types, such as single-qubit and two-qubit gates, each designed to perform specific transformations on quantum states. Quantum gates are reversible transformations, which means they allow for the exact reconstruction of the original input information after processing. When a quantum gate is applied to a set of qubits, the operation can be undone without any loss of information. Because quantum gates are reversible, they preserve the quantum information encoded in qubits.
    
    In quantum computing, quantum gates are represented by the unitary operators. Unitary operators are mathematical operators represented by matrices that satisfy the condition
    $$
        U^\dag U = I,
    $$
    where~$U^\dag$ is the adjoint (conjugate transpose) of~$U$, and~$I$ is the identity matrix.
    
    Quantum circuits are composed of sequences of quantum gates applied to qubits to perform specific computational tasks. Just as classical circuits are constructed from interconnected logic gates, quantum circuits are built by connecting quantum gates. They describe the flow of information and operations in the quantum computation.

    Within quantum circuits, interference emerges is a phenomenon where the probability amplitudes of different quantum states combine and interact. Transition amplitudes describe the probability amplitude for a qubit to transition from one quantum state to another under the influence of a quantum gate or operation. In quantum algorithms, transition amplitudes are manipulated by applying quantum gates to the quantum circuit. By carefully designing the sequence of gates, the interference effects can be exploited to enhance the probability of obtaining the desired output state while minimizing the probability of undesired outcomes. The interference can be constructive, where probability amplitudes increase the probability of a particular outcome, or destructive, where probability amplitudes cancel each other out, reducing the probability of specific outcomes. The ability to control transition amplitudes is a key feature that enables quantum computers to solve specific problems more efficiently than classical computers.

    State overlap and operator fidelity play a crucial role in quantifying the similarity between quantum states. State overlap quantifies the extent to which two quantum states share common elements or characteristics, providing insight into their similarity. Operator fidelity quantifies the accuracy of a quantum operation or transformation by measuring the closeness between the input and output states. Maximizing fidelity ensures the reliability and effectiveness of quantum algorithms, enhancing their computational performance and accuracy.

    Entanglement refers to a special relationship between qubits that allows them to become correlated in such a way that the state of one qubit directly influences the state of another, regardless of their individual locations within a quantum system. When two qubits are entangled, they form a single quantum state that cannot be described independently, which means that the measurement of one qubit will instantly determine the state of the other qubit, even if they are not physically connected. Entanglement enables quantum computers to perform calculations on multiple states simultaneously and exhibit non-local correlations, exponentially increasing processing power for certain problem domains.

    \subsection{Quantum Machine Learning}\label{subsec:2.2}
    Quantum machine learning explores the potential benefits of using quantum algorithms and quantum computing hardware to enhance classical machine learning tasks~\cite{NISQ2022}. We focus on enhancing the SVM algorithm, which is a widespread tool in the domain of machine learning-based malware detection~\cite{Ucci2019}, by combining it with a quantum kernel, estimated using a quantum computer.

    The quantum advantage lies in using a kernel, which is hard to estimate classically~\cite{Havlicek2019}. QSVM is based on quantum circuits that are hard to simulate due to their unique quantum properties, such as entanglement and superposition. QSVM promises to achieve better accuracy than conventional SVM across various problem domains, including malware detection~\cite{Barrue2023}.

    In this section, we explain the concept of kernels in SVM and introduce the quantum kernel. We also provide an overview of the tools used to implement and run software on quantum computers, specifically Qiskit and IBM Quantum. Additionally, we present related work in the field and discuss how it connects to our research.
    
    \subsubsection{QSVM}\label{subsubsec:2.2.1}
    In SVM classification, the algorithm seeks to find an optimal decision boundary that separates the data points into different classes. Once the decision boundary is established, new data points can be classified by determining which side of the boundary they fall on. Many real-world datasets are not inherently linearly separable, which is why kernels are used in SVM. Kernels map the input features to a new, possibly higher-dimensional space where the data may become more easily separable.

    A feature map~$\phi(x)$ is a function which maps each data point~$x$ from the original input feature space to a new transformed feature space with a higher dimensionality. The kernel function
    $$
        k(x,y) = (\phi(x) \cdot \phi(y))
    $$
    computes the dot product between two vectors~$x$ and~$y$ in the higher-dimensional feature space. Instead of explicitly computing the transformed vectors ~$\phi(x)$ and ~$\phi(y)$, the kernel function computes the dot product directly from the original input space without explicitly performing the mapping, which allows SVM to operate efficiently in high-dimensional space~\cite{Scholkopf1999}.

    There are various types of SVM kernels, such as polynomial, RBF, and sigmoid kernels. Different kernel functions define different ways of projecting the data and measuring similarity between points. QSVM combine SVM with a quantum kernel, computed using a quantum computer. The SVM model is then fitted with the precomputed quantum kernel and trained on a classical computer.

    The key difference between classical and quantum kernels lies in how the data are processed. In a classical kernel, the data are processed directly in the original form within the computational framework. The kernel computes the dot product between feature vectors in the original input space. This computation is done explicitly, without any transformation of the data into a different space.
    
    In contrast, the quantum kernel requires data to be transformed into quantum state space before processing. In the context of QSVM, we refer to this transformation as data encoding. Once the data are encoded, the quantum kernel function is applied to compute correlations between the quantum states. Therefore, estimating the quantum kernel involves two main components: the encoding of classical data and the application of the quantum kernel function.
    
    The data encoding process is done through a quantum feature map, denoted as~$\phi(x)$. It is a parameterized quantum circuit that maps a classical feature vector~$x$ to its corresponding quantum state~$\ket{\phi(x)}\bra{\phi(x)}$. The mapping is done by applying the unitary operation~$U_{\phi(x)}$ to the initial state~$\ket{0^n}$, where~$n$ is the number of qubits used for encoding ~\cite{Phan2021}. The index~$\phi(x)$ in the~$U_{\phi(x)}$ refers to the specific parameterization of the operation~$U$, which depends on the classical feature vector~$x$. Quantum gates and operations can be parameterized by certain variables, which affect how they transform quantum states. Different values of~$x$ lead to different parameterizations of the unitary operation, resulting in different quantum states after the transformation.

    The quantum kernel function 
    $$
        k(x,y) = \braket{\phi(x)}{\phi(y)} = \lvert \braket{\phi(x)}{\phi(y)} \rvert ^2
    $$
    is defined as the state overlap of the two data-encoded feature vectors from the quantum state space and represents the similarity between them~\cite{Havlicek2019}. A larger value of~$k(x,y)$ indicates that the classical data points~$x$ and~$y$ are close in feature space~\cite{Phan2021}.
    
    When applied to all datapoints, quantum kernel function generates the quantum kernel matrix
    $$
        K_{i,j} = k(x_i,x_j) = \lvert \braket{\phi(x_i)}{\phi(x_j)} \rvert ^2,
    $$
    where the entries represent the fidelities between different feature vectors. The fidelities can be computed efficiently on the quantum computer by calculating the transition amplitude between the states
    $$
        K_{i,j} = k(x_i,x_j) = \lvert \braket{\phi(x_i)}{\phi(x_j)} \rvert ^2 = \lvert \bra{0^n} U_{\phi(x_i)}^\dag U_{\phi(x_j)} \ket{0^n} \rvert ^2,
    $$
    where the feature map~$\phi(x)$ is described as the unitary operation~$U_{\phi(x)}$ applied to the initial state~$\ket{0^n}$~\cite{Havlicek2019, Phan2021}.

    \subsubsection{Qiskit}\label{subsubsec:2.2.2}
    In our work, we rely on Qiskit~\cite{qiskit2024} to implement the QSVM algorithm. Qiskit is an open-source software development kit for Python that enables users to design and implement algorithms for quantum computers at the level of quantum circuits. These algorithms can be executed locally on simulators or on quantum computers from IBM.
    
    IBM provides access to the quantum computers, known as IBM Quantum systems, via cloud through the IBM Quantum platform~\cite{IBMQuantumPlatform}, allowing researchers to experiment with real quantum hardware without needing specialized infrastructure. IBM processors fall under the NISQ devices category, meaning they operate with a limited number of qubits and suffer from errors due to the noise in the quantum hardware~\cite{NISQ2022, PreskillNISQ}. As of September 2024, eleven quantum processors are available on the IBM Quantum platform. Three quantum processors are freely available to the public, while the remainder is accessible via a premium plan.
    
    Qiskit Machine Learning~\cite{QiskitMachineLearning} is a module within Qiskit which provides tools for quantum machine learning tasks, including classical machine learning algorithms that can be enhanced using quantum computing techniques and entirely new quantum machine learning algorithms designed to run on quantum computers. We focus on introducing the classes implementing the quantum kernel within the Qiskit Machine Learning module. Understanding those classes is essential for effective integration of quantum-based kernels into the SVM.

    The quantum kernel interface is abstractly defined by the \texttt{BaseKernel}~\cite{BaseKernel} class. It specifies the \texttt{evaluate} method for constructing a kernel matrix from a given dataset, which is compatible with the Quantum Support Vector Classifier within Qiskit Machine Learning, as well as other kernel-based machine learning algorithms in established classical frameworks (for example, \texttt{scikit-learn}~\cite{scikit-learn}). Each entry in the kernel matrix is the result of the kernel function, defined as
    $$
        K(x,y) = \braket{f(x)}{f(y)},
    $$
    where~$x$,~$y$ are n-dimensional inputs and~$f$ is a map from an n-dimensional to an m-dimensional space. The quantum kernel algorithm computes the kernel matrix given the datapoints~$x$ and~$y$, and the feature map~$f$, all of dimension~$n$.

    The \texttt{FidelityQuantumKernel}~\cite{FidelityQuantumKernel} implements the \texttt{BaseKernel} interface. The kernel function is defined as the overlap of two quantum states~$x$ and~$y$,
    $$
        K(x,y) = \lvert \braket{\phi(x)}{\phi(y)} \rvert ^2,
    $$
    constructed using the feature map~$\phi(x)$. The \texttt{FidelityQuantumKernel} requires a fidelity primitive, which computes the fidelity between quantum states based on the \texttt{BaseStateFidelity}~\cite{BaseStateFidelity} algorithm introduced in Qiskit.
    
    The \texttt{BaseStateFidelity} class is an interface that calculates state fidelities (state overlaps) for pairs of (parameterized) quantum circuits. The fidelity calculation is generally defined as the state overlap
    $$
        \lvert \braket{\psi(x)}{\phi(x)} \rvert ^2,
    $$
    where~$\psi$ and~$\phi$ represent the states, and~$x$ and~$y$ are optional parameterizations of these states. The default fidelity primitive in the \texttt{FidelityQuantumKernel} is the \texttt{ComputeUncompute}~\cite{ComputeUncompute}, which implements the \texttt{BaseStateFidelity} interface.

    The data encoding process allows the quantum kernel to generate correlations between variables that are difficult to achieve using classical methods alone. The feature map must be based on quantum circuits that are hard to simulate classically~\cite{Havlicek2019} to obtain the quantum advantage over conventional kernels used in SVM. We describe feature maps based on the work of Havlicek et al.~\cite{Havlicek2019} and implemented in Qiskit, which we later use in our experiments, namely \texttt{PauliFeatureMap}~\cite{PauliFeatureMap}, \texttt{ZZFeatureMap}~\cite{ZZFeatureMap} and \texttt{ZFeatureMap}~\cite{ZFeatureMap}.
    
    The \texttt{PauliFeatureMap} is based on the Pauli matrices, which are fundamental operators in quantum mechanics. The Pauli matrices include the X, Y and Z matrices, each representing a different type of quantum operation. In the \texttt{PauliFeatureMap}, combinations of these matrices, specified by the \texttt{paulis} parameter, are applied to the input qubits to generate entanglement and capture features of the input data. The \texttt{PauliFeatureMap} typically consists of layers of single-qubit rotations and entangling gates involving Pauli matrices, with parameters that can be optimized during training to learn an adequate representation of the data for classification tasks. Data encoding is achieved by applying the unitary operation~$U_{\phi(x)}$ to the initial state, which in the case of \texttt{PauliFeatureMap} is defined as
    $$
        U_{\phi(x)} = \textrm{exp} \left( i \sum_{S \in \mathcal{I}} \phi_S(x) \prod_{i \in S} P_i \right),
    $$
    where~$S$ is a set of qubit indices that describes the connections in the feature map,~$\mathcal{I}$ is a set containing all these index sets,~$P_i$ refers to the chosen Pauli matrix, and
    $$
        \phi_S(x) =
            \begin{cases}
                x_i & \textrm{if } S = \left\{i\right\}\\
                \prod_{j \in S}(\pi - x_j) & \textrm{if } \lvert S \rvert > 1
            \end{cases}
   $$
    is the data mapping function, which can be customized.

    The \texttt{ZZFeatureMap} is a special case of the \texttt{PauliFeatureMap}, where the \texttt{ZZ} denotes the use of to the Pauli-Z matrices. These matrices represent the ZZ interaction between qubits, contributing to the entanglement in the quantum circuit. In the \texttt{ZZFeatureMap}, the Pauli matrices~$P_i$ are specifically chosen as Pauli-Z matrices, resulting in a product term that captures the ZZ interaction between qubits

    The \texttt{ZFeatureMap} is another specific case of the \texttt{PauliFeatureMap}. Unlike the \texttt{ZZFeatureMap}, it consists solely of Pauli Z matrices without entangling operations between qubits. As a result, the encoding produced by the \texttt{ZFeatureMap} does not exhibit entanglement. While this lack of entanglement may mean that the \texttt{ZFeatureMap} does not provide a quantum advantage for certain tasks, its effectiveness still depends on the specific problem being addressed.

    The last feature map we later use in our experiments is not implemented in Qiskit directly. However, it is based on the \texttt{ZZFeatureMap} with a custom custom data mapping function, defined as 
    $$
        \phi_S(x) =
            \begin{cases}
                x_i & \textrm{if } S = \left\{i\right\}\\
                \textrm{sin}(\pi - x_i)\textrm{sin}(\pi - x_j) & \textrm{if } S = \left\{i,j\right\},
            \end{cases}
    $$
    where \(S\) is a set of qubit indices that describes the connections in the feature map~\cite{Phan2021}. We later refer to this feature map as the \texttt{ZZphiFeatureMap}.

    All the feature maps mentioned can have a custom circuit depth specified by the \texttt{depth} parameter, which refers to the number of layers of quantum gates or operations applied to the input qubits to transform classical data into a quantum state. In the \texttt{PauliFeatureMap}, each layer typically consists of single-qubit rotations and entangling gates involving Pauli matrices. The depth of the \texttt{PauliFeatureMap} is determined by the number of these layers applied to the input qubits. The depth of a \texttt{PauliFeatureMap}, or any quantum circuit, represents the complexity of the circuit and the number of sequential operations used to encode classical data into a quantum state. A deeper circuit may capture more complex patterns in the data but may also require more computational resources.

    \subsection{Related Work}\label{subsec:2.3}
    The inspiration for our research is laid by the work of Barrué and Quertier~\cite{Barrue2023}, which provides insights into the performance of quantum machine learning algorithms in the context of malware detection. Notably, to date, this is the only paper that addresses malware detection through quantum computing methods while also performing experiments on IBM quantum computers rather than solely relying on simulators. Their work investigates QSVM alongside Quantum Neural Networks, and their findings underscore the potential of QSVM to outperform SVM with conventional kernels, mainly when operating with smaller datasets. Their research is heavily focused on experiments using only Qiskit's simulator. In contrast, our approach differs by concentrating on experiments with real quantum computers, which allows us to assess the practical challenges and performance of QSVM in a more realistic setting.
    
    However, we encountered several challenges when replicating their results due to the paper's lack of detailed experimental descriptions and parameter specifications. More importantly, they do not specify how many qubits and shots were used or which processors were utilized when conducting experiments on IBM Quantum devices. Additionally, they are not consistent with their metrics, such as not consistently measuring the F1-score, and if so, it is not clear to which parameters it belongs.
    

%% file: text/sections/3-section.tex
\section{Implementation}\label{sect:3}
    Our implementation consists of two main Python modules: the \texttt{peml} module, which is responsible for preprocessing the chosen dataset, and the \texttt{svm} module, which implements the SVM classification interface with both quantum and classical kernels. These modules are designed to function independently. The \texttt{peml} module focuses on preprocessing the specified dataset. The \texttt{svm} module can classify any dataset that adheres to the input format. The source code, along with detailed documentation, is available on GitLab~\cite{Kratka2024}.

    The \texttt{QSVM} class within the \texttt{svm} module implements the interface for QSVM classification using both the local simulator and quantum computers from IBM. Our implementation is based on two main classes from the Qiskit Machine Learning module~\cite{QiskitMachineLearning}, \texttt{ComputeUncompute}~\cite{ComputeUncompute} and \texttt{FidelityQuantumKernel}~\cite{FidelityQuantumKernel}, which we previously described in detail in Section~\ref{sect:2}. However, a significant limitation of these classes, and the Qiskit Machine Learning module as a whole, is that they are designed to run only on Qiskit's local quantum computer simulators. We aim to apply QSVM on real quantum hardware, specifically IBM's quantum computers.
    
    In this section, we explain the challenges encountered when running the code on actual quantum hardware and detail how and why we modified the source code of these two classes to overcome these obstacles, enabling execution on quantum devices. While the challenges are explained in the context of QSVM, they are universal to any large-scale practical quantum machine learning problem, not limited to QSVM, that requires substantial data processing on quantum hardware. For instance, similar issues would arise when implementing other models, such as neural networks.

    \subsection{Modifications for Quantum Hardware}\label{subsec:3.1}
    The implementation of the \texttt{ComputeUncompute} and \texttt{FidelityQuantumKernel} classes has three significant limitations that prevent the code from running on quantum hardware: inability to split the evaluation process, lack of transpilation for fidelity circuits and submission of all fidelity circuits in a single computational job, which exceeds the maximum job size limit. In the modified versions of the classes, we address these issues. Our improvements enable efficient resource utilization, ensure compatibility with IBM Quantum hardware, and enhance scalability for real-world machine learning applications.

    However, a major ongoing challenge is that Qiskit and its Qiskit IBM Runtime module constantly evolve, but often without maintaining minimal backward compatibility, which makes it difficult to keep the implementation up to date, and parts of the project may become outdated even in terms of few months. Nonetheless, as mentioned earlier, these three problems are not specific only to the QSVM implementation in Qiskit Machine Learning. They are general issues that need to be considered when working with real IBM Quantum hardware and should be accounted for in any project design.

    \subsubsection{Evaluation Process Must Wait for the Job Completion}\label{subsubsec:3.1.1}
    The original implementation of the classes lacks the ability to split the evaluation process into two distinct parts: submitting the computational jobs to IBM Quantum and processing the completed jobs. As a result, the classification process must run continuously while awaiting job execution on the IBM Quantum platform, which can take several days, depending on the job queue. This inefficiency not only consumes unnecessary resources but also restricts the scalability of the evaluation process, particularly when dealing with large datasets.

    To address this limitation, we introduced a solution that divides the process into two parts by adding helper methods to handle job submission and post-processing separately. In the first part, jobs are submitted to IBM Quantum to calculate the entries of the kernel matrix. Once the quantum jobs are completed, the second phase processes the results and evaluates the kernel matrices using the saved configuration.

    \subsubsection{Missing Transpilation}\label{subsubsec:3.1.2}
    Another issue is the absence of transpilation for the fidelity circuits before submitting computational jobs to IBM Quantum, which is a critical flaw in the original implementation. Transpilation refers to transforming quantum circuits to use only instructions supported by the targeted quantum processor. This transformation ensures compatibility and efficient execution on the actual quantum hardware. As of March 1, 2024, IBM Quantum introduced a significant update to improve the speed and efficiency of quantum computation~\cite{IBMQuantumDocsConfigureRuntimeCompilation, IBMQuantumPlatformUpdateQiskitPrimitives}. Circuits and observables must now be transformed to use only the instruction set architecture (ISA) supported by the target quantum system, meaning that all circuits must be transpiled before submission for execution.
    
    Without transpilation, the fidelity circuits in QSVM cannot be executed on IBM processors, which makes the \texttt{ComputeUncompute} and \texttt{FidelityQuantumKernel} classes unusable for real-world applications. It is worth noting that the transpilation issue is known and tracked by the Qiskit community, affecting several other classes beyond those discussed here, yet as of the completion of this work, it remains unresolved~\cite{QiskitAlgorithmsIssue164, QiskitIBMRuntimeIssue1519}.
    
    To address this issue, we added logic to ensure the fidelity circuits are transpiled before submission to the target quantum processor. However, while transpilation is necessary for executing quantum circuits on IBM hardware, it is not straightforward. It involves a series of optimizations that can sometimes alter the properties of the original circuit. During transpilation, circuits are transformed to match the constraints of the target system, such as available gates and qubit connectivity. However, this can result in issues such as increased circuit depth, which directly impacts execution time and noise levels.
    
    Additionally, circuits might be mapped to sub-optimal qubits for the specific computation, further degrading performance. In some cases, the original structure of the circuit, which was carefully designed for a specific behavior, may be lost or compromised during the transpilation process. These challenges make transpilation a problem of its own, requiring careful consideration when working with real quantum hardware, as the efficiency and accuracy of the quantum computation can be significantly affected.

    \subsubsection{Exceeding Maximum Job Size Limit}\label{subsubsec:3.1.3}
    The original classes submit all the fidelity circuits in a single computational job. While this approach works for local simulation, it becomes impractical for larger datasets on IBM Quantum systems. The job size often exceeds the maximum allowed limit~\cite{IBMQuantumDocsMaxExecutionTime}, preventing circuit execution and significantly limiting the usability of these classes, especially with larger datasets.
    
    To address this limitation, we implemented a one-job-per-kernel-entry approach, where each fidelity circuit responsible for computing one entry of the kernel matrix is computed in an individual job. We avoid unnecessary queuing delays by submitting these jobs in a session, enhancing overall efficiency and scalability.

%% file: text/sections/4-section.tex
\section{Experiments}\label{sect:4}
    This section describes the experiments we performed and presents our results. We categorize the experiments into two types: those run on Qiskit's local simulator and those executed on IBM Quantum processors. First, we outline the dataset and evaluation metrics used, followed by a detailed description of the experiments within each category.
    
    \subsection{Dataset}\label{subsec:4.1}
    We used the publicly available PE Malware Machine Learning Dataset~\cite{PEML} for our experiments. A key benefit of this dataset is that it provides the raw binary files themselves rather than just metadata extracted from the samples.

    The dataset consists of raw binaries of PE files, such as .exe or .dll files, and contains 201,549 labeled samples, with 86,812 benign and 114,737 malware samples. It is distributed in an encrypted zip folder, with file extensions removed from the individual samples to prevent accidental execution. Most malicious samples are sourced primarily from platforms like VirusShare~\cite{VirusShare}, MalShare~\cite{MalShare}, and theZoo~\cite{TheZoo}. Most of the legitimate files come from various instances of Windows 7, featuring a variety of installed software. However, there is a potential bias towards files associated with Microsoft products among them.

    Directly feeding raw binary files into the model is impractical due to their unstructured nature and the volume of data. Unstructured data lacks the organization and formatting necessary for practical analysis, and the amount of information in raw binary files makes it challenging for the model to extract meaningful patterns. Therefore we applied preprocessing techniques such as conversion to grayscale images~\cite{Nataraj2011} and Principal Component Analysis~\cite{PCA} to transform the raw binaries into informative feature vectors from which the model can learn.

    We converted the samples into grayscale images, adjusting their width based on the size of the binary content according to the predefined size ranges from Nataraj et al.\cite{Nataraj2011}. The images were resized to a uniform size while maintaining their aspect ratio and flattened into one-dimensional feature vectors. To align the dimensionality of the feature vectors with the number of qubits used in our experiments, we applied Principal Component Analysis (PCA) for dimensionality reduction. Although it may seem counterintuitive to convert binary files to images before applying PCA, we followed this approach to replicate the setup and results presented in the paper by Barrué and Quertier~\cite{Barrue2023}, described in Section~\ref{sect:2}. However, the image-construction process might not be necessary, and directly applying PCA to the binary data could have avoided the resizing and flattening steps. We randomly selected samples for the training and testing groups, ensuring an equal number of benign and malicious samples to create balanced datasets for our experiments.

    \subsection{Evaluation Metrics}\label{subsec:4.2}
    We adopt two metrics for evaluating the performance of models, accuracy and F1 score. Both metrics rely on the following terms, true positives, true negatives, false positives, and false negatives.
    \begin{itemize}
        \item{True positives (\textbf{TP}) refer to the number of malware samples that are correctly classified as malware.}
        \item{True negatives (\textbf{TN}) represent the number of benign samples correctly classified as benign.}
        \item{False positives (\textbf{FP}) refer to the number of benign samples that are incorrectly classified as malware.}
        \item{False negatives (\textbf{FN}) represent the number of malware samples that are incorrectly classified as benign (missed malware detections).}
    \end{itemize}
    
    Accuracy represents the proportion of correctly classified samples (both malware and benign) out of the total number of classifications~\cite{accuracyScore}. It provides a straightforward indication of the model's overall correctness, reflecting how often it gets the classification right.
    $$
        accuracy = \frac{\text{TP} + \text{TN}}{\text{TP} + \text{TN} + \text{FP} + \text{FN}}
    $$

    F1 score is defined as a harmonic mean of precision and recall~\cite{f1Score}. Precision measures how many of the samples classified as malware are truly malware~\cite{precisionScore}. For example, in malware detection, precision tells us what fraction of the files the model flagged as malware are actually malicious. Recall measures how many of the actual malware samples were correctly classified~\cite{recallScore}. It tells us how well the model performs in detecting malware. It indicates the proportion of all malware samples that the model successfully identifies.

    $$
        precision = \frac{\text{TP}}{\text{TP} + \text{FP}} 
    $$
    $$
        recall = \frac{ \text{TP} } { \text{TP} + \text{FN} }
    $$

    The F1 score combines precision and recall into a single metric, which can be especially useful when false positives and false negatives carry different consequences. In the context of malware detection, a high F1 score ensures that the model is not only accurate but also balances identifying actual malware and avoiding false positives, which can be critical when both false negatives (undetected malware) and false positives (benign files flagged as malware) are undesirable.
    $$
        \text{F1} = \frac{2}{\frac{1}{precision}\frac{1}{recall}} = \frac{2 \times \text{TP}}{2 \times \text{TP} + \text{FP} + \text{FN}}
    $$
    
    If there are no TP, FN, or FP samples (for example, in cases where no malware samples were predicted), the F1 score defaults to zero to avoid division errors. 
    
    \subsection{Experimental Results}\label{subsec:4.3} 
    Our primary focus was on testing and assessing performance on real quantum hardware. While simulators are flexible and convenient, they do not fully capture the complexity of quantum behavior under real conditions. They cannot fully emulate the effects of quantum noise in real quantum systems and come at a higher computational cost. However, testing the implementation first on a simulator is a crucial part of any quantum computing experiment. Simulators serve as a benchmark, helping to verify that the quantum circuit is correctly implemented.

    IBM Quantum computers offer the opportunity to validate algorithms under real-world conditions. Despite this advantage, running experiments on quantum computers introduces several challenges that affect the consistency and scalability of the results, as discussed in the previous section. Due to these limitations, the experiments conducted on IBM Quantum processors differ from those run on simulators. We could not run as many experiments on the hardware as on the simulator due to implementation constraints and limited access to computational resources.

    On both platforms, our goal was to evaluate the performance of our QSVM implementation and compare it with conventional SVM using kernels like polynomial or RBF. We focused primarily on the model's accuracy and investigated whether the QSVM demonstrated any quantum advantage in improved performance over classical methods.
    
    \subsubsection{Simulator}\label{subsubsec:4.3.1}
    On the local simulator in Qiskit, we tested QSVM classification with datasets of various sizes, ranging from 500 training samples and 100 test samples to 8000 training samples and 4000 test samples. For comparison, we performed SVM classification using classical kernels to evaluate how QSVM performs against conventional methods. Our goal was to replicate the experimental setup from  Barrué and Quertier~\cite{Barrue2023} as closely as possible and determine whether our implementation achieved similar performance improvements, particularly on smaller datasets.

    A notable finding from Barrué and Quertier~\cite{Barrue2023} is that quantum kernels, especially the \texttt{ZZFeatureMap}, demonstrated up to a 2.5\% improvement in accuracy over conventional SVM kernels in specific configurations. Their results suggest that QSVM may have an advantage in scenarios with limited dataset size. We aimed to verify these claims by comparing the performance of QSVM with classical SVM kernels across various dataset sizes.

    In the experiments, we used quantum kernels with different feature maps: \texttt{ZZFeatureMap} (ZZ), \texttt{PauliFeatureMap} (Pauli), \texttt{ZZphiFeatureMap} (ZZphi), and \texttt{ZFeatureMap} (Z), with the depth of the circuits set to 2. We used 1000 shots for all experiments, as the referenced paper did not specify the shot count. Number of shots refers to the number of repetitions of each circuit for sampling. Increasing the number of shots influences the statistical significance of the quantum measurements but at the cost of the computational time. Our input data consisted of binaries transformed into grayscale images of size 64×64, which were preprocessed into feature vectors of dimensions corresponding to the number of used qubits. The same preprocessing method was applied to both quantum and classical experiments, with the kernel being the primary differentiating factor.
    
    The results, presented in Table~\ref{tab:my_res_sim_full_accuracy} and Table~\ref{tab:my_res_sim_full_f1_score}, demonstrate that QSVM consistently matches or outperforms the accuracy and F1 scores of SVM using classical kernels.
    
    The results presented in Table~\ref{tab:my_res_sim_full_accuracy} and Table~\ref{tab:my_res_sim_full_f1_score} demonstrate that QSVM consistently matches or outperforms the accuracy and F1 scores of SVM using classical kernels. In Figure 1 and Figure 2, we highlight the performance of the kernels based on ZZ and ZZphi feature maps compared to the RBF kernel. Notably, the ZZ and ZZphi kernels exhibit the best performance among the quantum kernels, while the RBF kernel stands out among the classical ones.

\begin{figure}[!htb]
        \centering
        \includegraphics[width=
\textwidth]{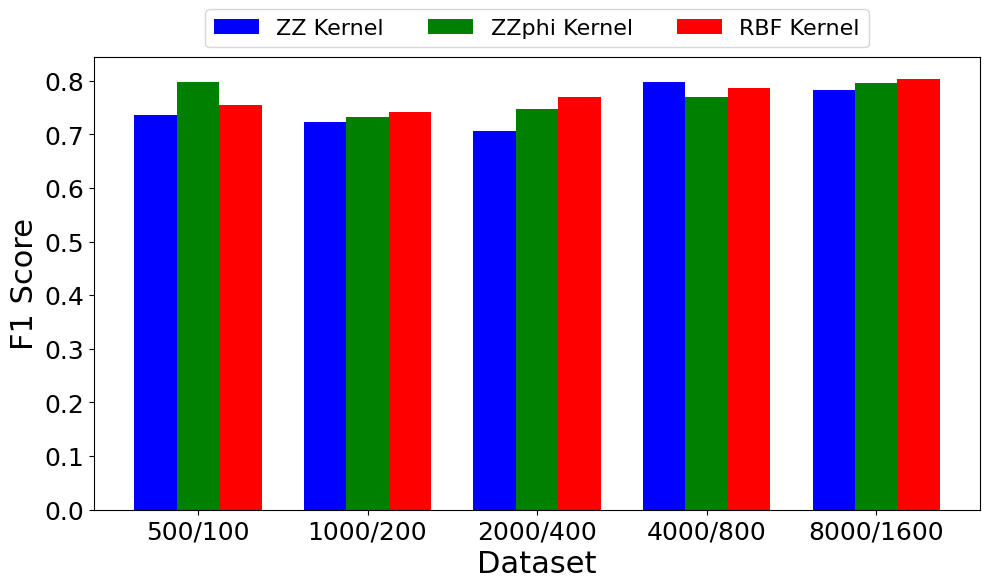}
        \caption{F1 Score Comparison With 3 Qubits}
        \label{fig:f1_score_3_qubits}
    \end{figure}

\begin{figure}[!htb]
        \centering
        \includegraphics[width=\textwidth]{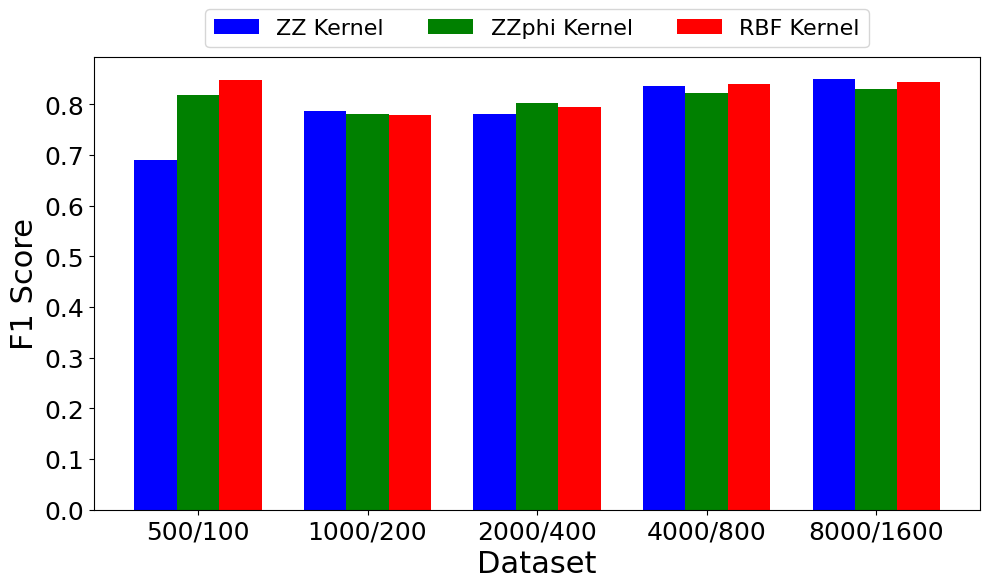}
        \caption{F1 Score Comparison With 7 Qubits}
        \label{fig:f1_score_7_qubits}
    \end{figure}

    Figure~\ref{fig:f1_score_3_qubits} displays the F1 score comparison for the ZZ, ZZphi, and RBF kernels with three qubits, illustrating how quantum approaches can remain competitive even with limited qubit resources. In contrast, Figure~\ref{fig:f1_score_7_qubits} presents the F1 score comparison for the same kernels with seven qubits, where the quantum kernels (particularly ZZ and ZZphi) achieve their highest F1 scores. Figure~\ref{fig:f1_score_7_qubits} provides a more comprehensive understanding of how these kernels scale with increased qubit count and data size, demonstrating the potential of quantum kernels against classical benchmarks like the RBF kernel.

        \begin{table}[!htb]
        \caption{Accuracy Comparison}
        \label{tab:my_res_sim_full_accuracy}
        \centering
        \begin{tabular}{l|l|l|l|l|l|l|l|l|l}
        \hline
            \multirow[b]{2}{*}{Data (Train/Test)} & \multirow[b]{2}{*}{Qubits} & \multicolumn{4}{l}{Quantum Kernels} & \multicolumn{4}{|l}{Classical Kernels} \\
            \cline{3-10} && ZZ & Pauli & ZZphi & Z & Linear & Polynomial & RBF & Sigmoid \\ \hline 
            500/100 & 3 & 0.740 & 0.790 & 0.800 & 0.780 & 0.750 & 0.690 & 0.760 & 0.510 \\ \hline
            ~ & 4 & 0.730 & 0.780 & 0.800 & 0.790 & 0.740 & 0.720 & 0.790 & 0.540\\ \hline
            ~ & 6 & 0.660 & 0.720 & 0.810 & 0.810 & 0.740 & 0.740 & 0.810 & 0.560\\ \hline
            ~ & 7 & 0.700 & 0.800 & 0.820 & 0.830 & 0.740 & 0.750 & 0.850 & 0.620\\ \hline
            
            1000/200 & 3 & 0.725 & 0.675 & 0.735 & 0.720 & 0.705 & 0.730 & 0.745 & 0.575\\ \hline
            ~ & 4 & 0.730 & 0.660 & 0.735 & 0.735 & 0.705 & 0.720 & 0.740 & 0.580\\ \hline
            ~ & 6 & 0.745 & 0.760 & 0.780 & 0.775 & 0.735 & 0.745 & 0.790 & 0.640\\ \hline
            ~ & 7 & 0.790 & 0.735 & 0.780 & 0.780 & 0.730 & 0.775 & 0.780 & 0.640\\ \hline
            
            2000/400 & 3 & 0.710 & 0.730 & 0.748 & 0.757 & 0.718 & 0.672 & 0.770 & 0.603\\ \hline
            ~ & 4 & 0.748 & 0.743 & 0.775 & 0.767 & 0.718 & 0.685 & 0.765 & 0.585\\ \hline
            ~ & 6 & 0.777 & 0.728 & 0.770 & 0.780 & 0.740 & 0.735 & 0.782 & 0.595\\ \hline
            ~ & 7 & 0.782 & 0.767 & 0.802 & 0.780 & 0.743 & 0.743 & 0.795 & 0.583\\ \hline

            4000/800 & 3 & 0.799 & 0.784 & 0.771 & 0.777 & 0.766 & 0.639 & 0.787 & 0.671\\ \hline
            ~ & 4 & 0.806 & 0.821 & 0.771 & 0.775 & 0.771 & 0.637 & 0.791 & 0.637\\ \hline
            ~ & 6 & 0.830 & 0.812 & 0.816 & 0.800 & 0.772 & 0.804 & 0.830 & 0.608\\ \hline
            ~ & 7 & 0.838 & 0.805 & 0.824 & 0.821 & 0.771 & 0.791 & 0.840 & 0.616\\ \hline
            
            8000/1600 & 3 & 0.783 & 0.779 & 0.797 & 0.796 & 0.779 & 0.619 & 0.804 & 0.633 \\ \hline
            ~ & 4 & 0.812 & 0.792 & 0.804 & 0.806 & 0.781 & 0.662 & 0.822 & 0.630\\ \hline
            ~ & 6 & 0.835 & 0.806 & 0.819 & 0.818 & 0.779 & 0.734 & 0.840 & 0.616\\ \hline
            ~ & 7 & 0.851 & 0.812 & 0.831 & 0.821 & 0.776 & 0.746 & 0.845 & 0.608 \\ \hline
        \end{tabular}
    \end{table}
    
    \begin{table}[!htb]
        \caption{F1 Score Comparison}
        \label{tab:my_res_sim_full_f1_score}
        \centering
        \begin{tabular}{l|l|l|l|l|l|l|l|l|l}
        \hline 
            \multirow[b]{2}{*}{Data (Train/Test)} & \multirow[b]{2}{*}{Qubits} & \multicolumn{4}{l}{Quantum Kernels} & \multicolumn{4}{|l}{Classical Kernels} \\
            \cline{3-10} && ZZ & Pauli & ZZphi & Z & Linear & Polynomial & RBF & Sigmoid \\ \hline 
            500/100 & 3 & 0.736 & 0.790 & 0.797 & 0.777 & 0.746 & 0.662 & 0.754 & 0.510\\ \hline
            ~ & 4 & 0.729 & 0.779 & 0.797 & 0.788 & 0.736 & 0.700 & 0.787 & 0.540\\ \hline
            ~ & 6 & 0.649 & 0.716 & 0.808 & 0.810 & 0.736 & 0.729 & 0.808 & 0.560\\ \hline
            ~ & 7 & 0.690 & 0.795 & 0.819 & 0.829 & 0.736 & 0.738 & 0.849 & 0.620\\ \hline

            1000/200 & 3 & 0.723 & 0.675 & 0.732 & 0.717 & 0.700 & 0.728 & 0.742 & 0.574\\ \hline
            ~ & 4 & 0.729 & 0.658 & 0.732 & 0.731 & 0.700 & 0.719 & 0.737 & 0.579\\ \hline
            ~ & 6 & 0.744 & 0.756 & 0.779 & 0.775 & 0.730 & 0.738 & 0.789 & 0.640\\ \hline
            ~ & 7 & 0.787 & 0.731 & 0.779 & 0.780 & 0.725 & 0.771 & 0.779 & 0.640\\ \hline

            2000/400 & 3 & 0.707 & 0.728 & 0.747 & 0.756 & 0.716 & 0.651 & 0.769 & 0.601\\ \hline
            ~ & 4 & 0.746 & 0.741 & 0.774 & 0.767 & 0.716 & 0.670 & 0.764 & 0.584\\ \hline
            ~ & 6 & 0.776 & 0.726 & 0.769 & 0.780 & 0.739 & 0.729 & 0.782 & 0.595\\ \hline
            ~ & 7 & 0.781 & 0.764 & 0.802 & 0.780 & 0.742 & 0.737 & 0.795 & 0.582\\ \hline
            
            4000/800 & 3 & 0.797 & 0.783 & 0.769 & 0.775 & 0.764 & 0.612 & 0.786 & 0.671\\ \hline
            ~ & 4 & 0.805 & 0.821 & 0.770 & 0.773 & 0.769 & 0.621 & 0.790 & 0.637\\ \hline
            ~ & 6 & 0.830 & 0.811 & 0.816 & 0.799 & 0.771 & 0.803 & 0.830 & 0.607\\ \hline
            ~ & 7 & 0.837 & 0.803 & 0.823 & 0.821 & 0.769 & 0.790 & 0.840 & 0.616\\ \hline
        
            8000/1600 & 3 & 0.783 & 0.779 & 0.796 & 0.794 & 0.778 & 0.589 & 0.804 & 0.633\\ \hline
            ~ & 4 & 0.812 & 0.792 & 0.803 & 0.805 & 0.779 & 0.646 & 0.822 & 0.630\\ \hline
            ~ & 6 & 0.835 & 0.806 & 0.818 & 0.818 & 0.778 & 0.731 & 0.840 & 0.616\\ \hline
            ~ & 7 & 0.851 & 0.812 & 0.830 & 0.821 & 0.775 & 0.743 & 0.845 & 0.607\\ \hline
        \end{tabular}
    \end{table}

    \subsubsection{IBM Quantum Systems}\label{subsubsec:4.3.2}
    The second phase of our experiments involves QSVM classification using quantum kernels computed on IBM Quantum computers, to which we have access thanks to a license from the Czech Technical University in Prague (CTU).
    
    Inspired by the potential of NISQ computers, our initial goal was to implement and evaluate QSVM primarily on IBM Quantum computers. However, during implementation, we encountered several challenges that significantly altered the course of our experiments, as detailed in Section~\ref{sect:3}. These challenges stem mainly from limitations within the Qiskit Machine Learning module, particularly regarding transpilation requirements and constraints on job sizes when using IBM Quantum systems.
    
    As a result, we faced limitations when running experiments on real quantum hardware. To mitigate these issues, we implemented a fix involving the addition of transpilation and adopting a one-job-per-kernel-entry approach, as described in Section~\ref{sect:3}. Transpilation, a critical requirement for executing quantum circuits on IBM Quantum systems, involves adapting circuits to conform to the target quantum system's ISA. While our fix addressed the critical obstacles, more efficient and optimal solutions likely exist. Unfortunately, due to time constraints during the project, we were unable to fully explore these alternatives. As a result, we were limited to testing small datasets, with a maximum of 20 training samples and 10 test samples.

    QSVM classification requires two quantum kernel matrices: one for training and one for testing. The training matrix is symmetric and has a size of ~$n \times n$, where~$n$ is the number of training samples. The test matrix is~$m \times n$, where~$m$ is the number of testing samples. For the dataset of 20 training and 10 testing samples, our one-job-per-kernel-entry approach results in 390 jobs on the quantum computer.
    
    During the debugging phase, we conducted experiments to evaluate the time required to execute a single job. Although the individual jobs are relatively small regarding data volume and processing time, the nature of machine learning tasks requires a substantial number of jobs, particularly with our current implementation, where one job is required per kernel entry. Each job involves running a parameterized quantum circuit (based on the chosen feature map) with a specific sample (feature vector) as the parameter. We tested different numbers of shots and various quantum processors, finding that executing one job takes approximately 15 seconds of \textit{quantum time}. Quantum time refers to the total duration a quantum system is committed to fulfilling a user's request~\cite{quantumTime}. Therefore, the total time required to evaluate the small dataset with 20 training and 10 testing samples is approximately 97.5 minutes on the quantum computer. These limitations are further compounded by the constraints of the CTU license, which grants us access to only 400 minutes per month.
    
    We experimented with the number of jobs submitted in a single session. Sessions allow all jobs to be executed consecutively, minimizing queue wait times. However, as the number of jobs and the quantum minutes used approach the limits imposed by our license, queue wait times can increase exponentially. Consequently, even small datasets (e.g., 20 train and 10 test samples) could queue for up to approximately 14 days on the \texttt{ibm\_torino} system, leading us to explore alternative systems.
    
    In our experimentation, we tested various systems and opted to submit all jobs within a single session to manage larger workloads. When selecting the least busy system available, we typically encountered queue times of only a few minutes. However, with the busiest systems (in our case, \texttt{ibm\_torino}), wait times could extend to several hours, even for a relatively small number of jobs. While the quantum processing time required to execute the jobs was consistent across various systems, with differences of only a few seconds, these variations had a notable impact given our limited resources and the larger volume of jobs we needed to process.
    
    Table~\ref{tab:20_10_hardware} presents the results of our experiments. We used different IBM Quantum systems for each dataset size, including \texttt{ibm\_torino}, \texttt{ibm\_algiers}, \texttt{ibm\_cairo}, and \texttt{ibm\_kyoto}. The column labeled \textit{job time} specifies the average execution time of each job on the respective system, measured in \text{quantum seconds}. Although we provide accuracy and F1 score metrics for completeness, it is important to note that due to the small dataset sizes, these metrics may not fully represent the performance of the QSVM algorithm. However, they offer insights into relative performance across different systems and dataset sizes. The table highlights the iterative nature of our experiments, beginning with smaller datasets and progressively scaling up. We started with 4 training and 2 test samples, gradually increasing to 8 training and 4 test samples, and eventually evaluating a larger dataset with 20 training and 10 test samples.
    
    \begin{table}[!htb]
        \caption{~Experiment Results: QSVM Classification on IBM Quantum Systems}
        \label{tab:20_10_hardware}
        \centering
        \begin{tabular}{l|l|l|l|l}
            \hline
            Data (Train/Test) & IBM Quantum System & Job Time & Accuracy & F1 Score \\ \hline
            \multirow[b]{2}{*}{4/2} & \texttt{ibm\_torino} & 15s & 0.5 & 0.333 \\ \cline{2-5}
            & \texttt{ibm\_algiers} & 18s & 0.5 & 0.333 \\ \hline
            \multirow[b]{2}{*}{8/4} & \texttt{ibm\_torino} & 18s & 1 & 1 \\ \cline{2-5}
             & \texttt{ibm\_algiers} & 15s & 0.75 & 0.733 \\ \hline
            \multirow[b]{2}{*}{20/10} & \texttt{ibm\_cairo} & 16s & 0.6 & 0.6 \\ \cline{2-5}
             & \texttt{ibm\_kyoto} & 17s & 0.6 & 0.524 \\ \hline
        \end{tabular}
    \end{table}

%% file: text/sections/5-section.tex
\section{Conclusion and Future Work}\label{sect:5}
    We extended the previous work by focusing on the implementation and evaluation on real quantum computers, which brings its own challenges. We addressed and fixed the issues in the original implementation of classes for quantum kernel in Qiskit Machine Learning library, namely the inability to split the evaluation process into distinct parts, the absence of transpilation for fidelity circuits and the issue with submitting all the fidelity circuits in one single job to IBM Quantum leading to exceeding the maximum limit for job size. The absence of transpilation is a known issue within the Qiskit community and has not yet been resolved at the time of finishing this work. Our fixes address critical flaws in the original implementation and pave the way for more efficient usage of quantum computing resources in malware detection.
    
    Besides the local simulator, we also used IBM Quantum computers to compute the quantum kernel for QSVM classification. We tested how the IBM Quantum computers behave under the workload of many computation jobs.
    
     In future work, we aim to optimize the transpilation process and the one-job-per-kernel entry approach to enable large-scale experiments on IBM Quantum computers. Further investigation into their topology would also be beneficial, as each system features a unique layout of qubits. We may reduce computation time by specifying the exact qubits used for computation.
    
    From an algorithmic perspective, we plan to experiment with feature map design and combine different data mapping functions to enhance our approach. Furthermore, we would like to investigate various preprocessing techniques and their impact on the classification results.